\documentstyle[11pt]{article}

\newcommand{\ppp}{{\partial}}
\newcommand{\ggg}{{4\pi G}}
\newcommand{\gggg}{{2\pi G}}

\renewcommand{\theequation}{\arabic{section}.\arabic{equation}}
\newcommand{\newsection}[1]{
\pagebreak[3]
\setcounter{equation}{0}
\renewcommand{\theequation}{\arabic{section}.\arabic{equation}}
\section{#1}
\nopagebreak
\medskip
\nopagebreak}

\begin{document}
\font\ninerm = cmr9
\baselineskip 14pt plus .5pt minus .5pt
\def\footnoterule{\kern-3pt \hrule width \hsize \kern2.6pt}
\hsize=6.0truein
\vsize=9.0truein
\textheight 8.5truein
\textwidth 5.5truein
\voffset=-.4in
\hoffset=-.4in
\pagestyle{empty}
\begin{center}
{\large\bf String-Inspired Gravity with Interacting
Point Particles}\footnote{\ninerm
\hsize=6.0truein
This work is
supported in part by funds provided by the U.S.
Department of Energy (D.O.E.)
under contracts \#DE-AC02-76ER03069 and \#DE-AC02-89ER40509,
as well as in part
by the National Science Foundation under contracts \#INT-910559 and
\#INT-910653.}
\vskip 1cm
Dongsu Bak and Domenico Seminara
\vskip 0.5cm
{\it Center for Theoretical Physics\\
Laboratory for  Nuclear Science and Department of Physics\\
Massachusetts Institute of Technology\\
Cambridge, Massachusetts 02139, U.S.A.}

\end{center}
\baselineskip=24pt
\vspace{1.2cm}
\begin{center}
{\bf ABSTRACT}
\end{center}
We reformulate  two dimensional string-inspired
gravity with point particles as a gauge
theory of the extended Poincar\'e group. A  non-minimal
gauge coupling is necessary for the equivalence of the two descriptions. The
classical one-particle problem is analyzed completely.  In addition, we obtain
the  many-particle effective action after eliminating the gravity
degrees of freedom. We investigate properties of this effective action, and
show how to recover the geometrical description.
Quantization
of the gauge-theoretic model is carried out and the explicit
one-particle solution is found.  However,
we show that the formulation leads to a quantum mechanical
inconsistency in the two-particle case. Possible cures are discussed.
\bigskip
\bigskip
\bigskip
\begin{center}
Submitted to: PRD 15
\end{center}
\bigskip
\bigskip
\vfill
CTP \#2431
\hfill
May 1995
\eject
\pagestyle{plain}
\pagenumbering{arabic}
\setcounter{page}{1}
\newsection{Introduction}

There have been many attempts to describe gravity in terms of
a gauge theory\cite{Poin}, since gauge theories enjoy a
well-defined  quantization procedure.
These include the Ashtekar formulation of Einstein gravity\cite{Asht},
Poincar\'e gravity\cite{Gri1}, Chern-Simons gravity\cite{Witt}
and so on. Recently,  string-inspired gravity\cite{CGHS} was also
reformulated as a gauge theory\cite{Ver1} using  the extended
Poincar\'e  group\cite{Can92}. Its quantization
was carried out and it was  found that the quantum solution of pure
gravity is described  by a rather trivial state characterized by
two constant modes of the gravity fields --- constants that can be
interpreted as the black hole mass and the cosmological constant\cite{Can92}.
The quantum equivalence to the geometric approach in \cite{kun} has been
proven by a direct comparison of wave functionals\cite{Ben}.
A model in which point particles couple to gravity in a gauge invariant
fashion, is  available as well\cite{Can93}. Its explicit many-particle
quantum solutions has been found. One unsatisfactory aspect of the solution is
that
the particles do not feel any gravitational interactions, even though
gravity fields feel the presence of particles. The reason
follows from the fact that in the model\cite{Can93},  particles couple to
the gauge metric that is flat and differs from the original physical
metric by a conformal factor,  while the particles are massive and
therefore not conformally invariant.

In this paper, we introduce a gauge-theoretic model, in which massive
particles couple to the physical metric of the string-inspired gravity.
In order to provide a gauge invariant description for matter particles,
one introduces Poincar\'e coordinates as additional dynamical variables.
We shall show that
the model is classically equivalent to the geometric formulation of
the string-inspired gravity
with point particles
by comparing
equations of motion in a specific gauge called the ``unitary'' gauge.
We  solve the classical one-particle problem and discuss the equivalence
at the level of the solution space.
We also analyze the classical many-particle problem, and derive an effective
Lagrangian for particles in which gravity variables are completely
eliminated.

When pursuing quantum solutions of the model, one
of our  main concerns is  whether the enlargement
of dynamical variables causes any problems in the
equivalence between the geometric approach and gauge-theoretic formulation.
We shall see that in the present model, inconsistencies at the quantum level
do arise from the enlargement for the case of more than one particle.
We shall clarify the origin of the inconsistencies and discuss possible
cures in detail.

\newsection{Gauge formulation of lineal gravity}

\noindent
When the string-inspired gravity was first proposed\cite{CGHS}, its geometrical
action was taken to be
\begin{equation}
\label{GG}
I=\frac{1}{4\pi G}\int d^2 x \sqrt{- g_{_P}} e^{-2\phi} (
R(g_{_P})+
4 g_{_P}^{\mu\nu}\partial_\mu \phi\partial_\nu \phi -\lambda).
\end{equation}
where $\lambda$ is the cosmological constant and $\phi$  the
dilaton field.\footnote{\hsize=6.0truein
Notation: the signature of  the metric
tensor $g^{_P}_{\mu\nu}$ is assumed
to be $(1,-1)$. The Latin indices ${a,~b,~c\dots}$ run over a tangent space
where the flat Minkowski metric $h_{ab}={\rm diag}(1,-1)$ is defined.
The antisymmetric symbol $\epsilon^{ab}$ is normalized by $\epsilon^{01}=1$.}
   Subsequently it was recognized that the
introduction of the new variables
\begin{equation}
\label{eq2}
g_{\mu\nu}=e^{-2\phi}  g^{_P}_{\mu\nu}
\ \ \ \ {\rm and}\ \ \ \ \eta=e^{-2\phi}
\end{equation}
transforms the action (\ref{GG}) into a simpler expression
\begin{equation}
\label{FF}
I=\frac{1}{4\pi G}\int d^2 x \sqrt{-g} (\eta R(g) -\lambda),
\end{equation}
which can be reformulated as a gauge theory\cite{Ver1,Can92,Gri94}. In fact
a gauge theoretical formulation of the action (\ref{FF})
can be given by using the 4-parameter extended Poincar\'e
group in 1+1 dimensions \cite{Can92,Can93,Can94},  with the
Lie algebra
\begin{eqnarray}
\label{algebra}
&&[P_a,P_b]=\epsilon_{ab} I\,,\ \ \ \ \ \ \ [P_a,J]=\epsilon_{a}^{~b} P_b \\
&&[P_a,I]=[J,I]=0\,,\nonumber
\end{eqnarray}
where  $I$ is the central element that modifies the  conventional algebra
of translation generators $P_a$, while  the Lorentz boost sector is
 unchanged. The extension arises naturally in two dimensions
if one  allows non-minimal gravitational coupling, as pointed
out in Ref. \cite{Can92}.

The gauge field is now introduced as a connection one-form that takes
values in the Lie algebra
\begin{equation}
\label{connection}
A_\mu=e^a_\mu P_a +\omega_\mu J+a_\mu I,
\end{equation}
where $e^a$ and $\omega$ are, respectively, the Zweibein and the
spin connection. The potential $a_\mu$ is
related to the volume form\cite{Can92}. The
connection defined in (\ref{connection}) transforms according to
the adjoint representation. In components the transformation reads
\begin{eqnarray}
&&(e^U)^a_\mu=(\Lambda^{-1})^a_{~b}(e^b_\mu+\epsilon^b_{~c}
\theta^c\omega_\mu
+\partial_\mu\theta^b)\\
&&(\omega^U)_\mu=\omega_\mu +\partial_\mu \alpha\\
&&(a^U)_\mu=a_\mu-\theta^a\epsilon_{ab} e^b_\mu-
\frac{1}{2}\theta^a\theta_a \omega_\mu+\partial_\mu\beta
+\frac{1}{2}\partial_\mu\theta^a\epsilon_{ab}\theta^b
\end{eqnarray}
where we have parameterized the gauge transformation as follows
\begin{equation}
U={\rm exp}(\theta^a P_a){\rm exp}(\alpha J){\rm exp}(\beta I)
\label{gaugep}
\end{equation}
and $\Lambda^a_{~b}$ is the Lorentz transformation matrix
\begin{equation}
\Lambda^a_{~b}=\delta^a_{~b} \cosh\alpha+\epsilon^a_{~b}\sinh\alpha.
\end{equation}
\noindent
The field strength associated to the connection (\ref{connection})
is now computed from
its definition
\begin{eqnarray}
\label{curvature}
&&F=dA+[A,A]\nonumber\\
&&=(de^a +\epsilon^a_{~b}\omega\wedge e^b) P_a+ d\omega J+ (da+
\frac{1}{2}\epsilon_{ab} e^a\wedge e^b) I.
\end{eqnarray}
To construct an invariant action linear in the curvature,
we  introduce a multiplet, $\eta_A\equiv (\eta_a,\eta_2,\eta_3)$,
that transforms according to the co-adjoint representation
\begin{eqnarray}
&&(\eta^U)_a=(\eta_b-\eta_3\epsilon_{bc}\theta^c)\Lambda^b_{~a}\\
&&(\eta^U)_2=\eta_2-\eta_a\epsilon^a_{~b}\theta^b+
\frac{1}{2}\eta_3 \theta^a\theta_a\\
&&(\eta^U)_3=\eta_3\,.
\end{eqnarray}
Note that $\eta_a$ may be set to zero by a gauge transformation.
The  action is now simply formed by contracting   $\eta_A$ with
$\epsilon^{\mu\nu} F^A_{\mu\nu}$
\begin{equation}
I_g=\frac{1}{4\pi G}\int d^2 x \epsilon^{\mu\nu}\Bigl(\eta_a D_\mu e^a_\nu+
\eta_2 \partial_\mu\omega_\nu+\eta_3 (\partial_\mu a_\nu+\frac{1}{2}
\epsilon_{ab} e^a_\mu e^b_\nu)\Bigr)\,.
\end{equation}
It is easy to show \cite{Can92,Can93} that this {\it B-F} theory is
equivalent to the string-inspired gravity defined by the geometrical action
(\ref{FF}) once  we identify $\eta_2=2 \eta $ and $ g_{\mu\nu}=
e^a_\mu e_{a\nu}$. The cosmological constant, $\lambda$, is generated
dynamically by the field $\eta_3$, which is fixed to be a constant by
the equations of motion.

A gauge invariant description of the  matter requires an
introduction of a new variable, the Poincar\'e coordinate $q^a$.
In particular,  for the case of point particles, a possible first-order
invariant action  is \cite{Gri1,Can94}
\begin{eqnarray}
\label{action1}
I_m=\int d\tau \Bigl [ p_a (D_\tau q)^a-\frac{1}{2} N(p^a p_a+m^2)
\Bigr ]     \\
(D_\tau q)^a\equiv \dot q^a+\epsilon^a_{~b}
\Bigl(q^b \omega_\mu(X)-e^b_\mu(X)\Bigr)\dot
X^\mu,
\end{eqnarray}
where $\tau$ is the trajectory parameter and the dot denotes
derivative with respect to $\tau$. [To avoid a cumbersome notation,
we just use $X$ when $X^\mu(\tau)$ appears as argument of a function.]
The particle is described by
the dynamical variables $q^a(\tau)$, $p^a(\tau)$ and $X^\mu(\tau)$,
while $N(\tau)$ is a Lagrange multiplier that enforces the mass-shell
condition.  Under a gauge  transformation  the gauge variables,
$q^a(\tau)$ and  $p^a(\tau)$ transform  according to the following rule
\begin{eqnarray}
\label{pp}
&&(q^U)^a=(\Lambda^{-1})^a_{~b}(q^b(\tau) +
\epsilon^b_{~c}\theta^c(X))\\
\label{qq}
&&(p^U)^a=\Lambda^a_{~b} p^b(\tau)\,,
\end{eqnarray}
where all the  gauge parameters are computed along the trajectory
$X^\mu$.

According to (\ref{pp}) the gauge, $q^a=0$, is always available; when it is
chosen, the usual geometrical action for a point particle
\begin{equation}
\label{jj}
I_m=\frac{1}{2}\int d\tau
\Bigl ( \frac{\dot X^\mu  g_{\mu\nu}(X)
\dot X^\nu}{N}-N m^2\Bigr )
\end{equation}
is recovered by eliminating $p^a$. (In a certain sense
the $q^a$ field looks like a Higgs field in a gauge theory with symmetry
breaking: its presence insures the gauge invariance, but when
the unitary gauge,  $q^a=0$, is chosen the physical
content of the theory is exposed.)

Though the action in (\ref{action1}) appears
natural in the gauge-theoretic approach,  the particles
do not follow  geodesics of the physical metric $g^{_P}_{\mu\nu}$.
Instead, $I_m$ describes particles moving along  geodesics of the gauge metric
that differ from the physical one by the conformal factor $\eta$. Since the
particles are massive, such a conformal redefinition changes the dynamics.
To make the particles move in the physical metric, a suitable modification
of the action
(\ref{action1}) is needed. In particular, an  introduction of a non-minimal
coupling with the $\eta^A$ fields is required.

\newsection{The model and its classical solutions}

In this section we shall investigate the string-inspired  gravity
coupled to  point particles. In the geometric formulation, the
action\cite{CGHS}  is given by
\begin{eqnarray}
\label{eq1}
I&=&\frac{1}{4\pi G}\int d^2 x \sqrt{-g_{\!_P}}
e^{-2\phi} (R(g_{\!_P})+
4 g_{\!_P}^{\mu\nu}\partial_\mu \phi\partial_\nu \phi -\lambda)\nonumber\\
&+&\frac{1}{2}\int d\tau \Bigl( \frac{\dot X^\mu g^{_P}_{\mu\nu}(X)\dot X^\nu}
{N}
 -N m^2\Bigr)
\end{eqnarray}
where $X^\mu(\tau)$ is the particle trajectory and $m$ is the mass.
In order to connect the geometrical approach with the gauge formulation
of the previous section, we introduce the new variables defined in
Eq.~(\ref{eq2}). The action (\ref{eq1}) now reads
\begin{equation}
\label{eq4}
I=\frac{1}{4\pi G}\int d^2 x \sqrt{- g} (\eta R -\lambda)
- \frac{1}{2} \int d\tau
\Bigl ( \frac{\dot X^\mu  g_{\mu\nu}(X)
\dot X^\nu}{N\eta(X)}-N m^2\Bigr ).
\end{equation}
The gravity sector takes  the same form of  the gauge action in (\ref{FF}).
The particle is  coupled not only to the gauge metric
$g_{\mu\nu}$, but also to the  dilaton field $\eta$. This explains
why the action  (\ref{action1}) is not suitable to describe
the matter sector of the action in (\ref{eq1}).\footnote{\hsize=6.0truein
We notice
that if we consider
a particle with zero mass,
the dependence on $\eta$ can be eliminated
through a redefinition  of the Lagrange  multiplier $N(\tau)$.
This is related to the fact that the light-like
geodesics are preserved by  conformal transformations of the metric.}

A way of overcoming this difficulty is to consider the following gauge
invariant action for the particle as well as the gravity
\begin{eqnarray}
\label{action2}
I&=&\frac{1}{4\pi G}\int d^2 x \epsilon^{\mu\nu}\Bigl(\eta_a D_\mu e^a_\nu+
\eta_2 \partial_\mu\omega_\nu+\eta_3 (\partial_\mu a_\nu+\frac{1}{2}
\epsilon_{ab} e^a_\mu e^b_\nu)\Bigr)\nonumber\\
&+&\int d\tau \Bigl [ p_a (D_\tau q)^a-\frac{1}{2} N
\Bigl(\frac{q^A\eta_A(X) }{2}p^2
+m^2\Bigr) \Bigr ]
\end{eqnarray}
where $q^A\eta_A (X)$ is the gauge invariant combination $\eta_a q^a +\eta_2
+\frac{1}{2} \eta_3 q_a q^a $ computed along the trajectory $X^\mu$.

The equivalence between the two actions can be easily shown by comparing
all the equations of motion in
$q^a=0$ gauge
with those in the geometric formulation.\footnote{\hsize=6.0truein
This
equivalence is more transparent if we
rewrite the particle action in a second order formalism
$$
I_m=\int d\tau
\frac{1}{2}\left ( \frac{D_\tau q^a D_\tau q_a}{N q^A\eta_A(X)}
-N m^2\right ).
$$
In fact, for $q^a=0$ this action collapses immediately to the geometrical
action. However, a reliable proof requires checking the equivalence of all
the equations of motion.
}

Let us now write down the equations of motion  and look for
the one body solution.  For the gravity sector,
we have
\begin{eqnarray}
\label{etaa}
&&\delta\eta_a\rightarrow S^a_{\mu\nu}=-\pi G \epsilon_{\mu\nu}
\int d\tau N p^2q^a
\delta^2 (x-X)\\
\label{eta2}
&&\delta\eta_2\rightarrow R_{\mu\nu}=-\pi G\epsilon_{\mu\nu}
\int d\tau N p^2
\delta^2 (x-X)\\
\label{eta3}
&&\delta\eta_3\rightarrow  \epsilon^{\mu\nu} (\partial_\mu a_\nu +\frac{1}{2}
\epsilon_{ab} e^a_\mu e^b_\nu)={\pi G\over 2}\int d\tau  N
p^2 q^2 \delta^2 (x-X)\\
\label{ea}
&&\delta e^a\rightarrow \partial_\mu \eta_a+\epsilon_{ab} \eta^b\omega_\mu
+\epsilon_{ab} e^b_\mu\eta_3=-4 \pi G \epsilon_{ab}\epsilon_{\mu\nu}
\int d\tau p^b \dot X^\nu \delta^2 (x-X)\\
\label{omega}
&&\delta\omega_\mu \rightarrow \partial_\mu \eta_2+\epsilon_{ab}\eta^a e^b_\mu
=-4 \pi G \epsilon_{ab}\epsilon_{\mu\nu}
\int d\tau p^a q^b \dot  X^\nu \delta^2 (x-X)\\
\label{a}
&&\delta a_\mu\rightarrow \partial_\mu \eta_3=0,
\end{eqnarray}
Here $\delta^2(x-X)$ stands for $\delta(x^0-X^0(\tau))
\delta(x^1-X^1(\tau))$, \  $S^a_{\mu\nu}(\equiv D_\mu e^a_\nu-D_\nu e^a_\mu)
$ denotes the torsion two-form and $R_{\mu\nu}(\equiv \partial_\mu\omega_\nu-
\partial_\nu\omega_\mu)$ is the curvature two-form in terms
of the spin connection. The matter sector produces three independent
 equations
\begin{eqnarray}
\label{N}
&&\delta N\rightarrow  \frac{\eta_A q^A}{2}p^2 +m^2=0\\
\label{p}
&&\delta p^a\rightarrow D_\tau q^a -N p^a{\eta_A q^A\over 2}=0\\
\label{q}
&&\delta q^a \rightarrow D_\tau p^a -\frac{N p^2}{4}(\eta^a +q^a\eta_3)=0
\end{eqnarray}
where $D_\tau p^a\equiv \dot p^a+\epsilon^a_{~b} p^b \omega_\mu \dot X^\mu$.
The variation with respect to $X^\mu$ merely produces an equation of
motion, which can be expressed as a linear
combination of Eqs.~(\ref{etaa}-\ref{q}). This linear dependence
of the particle equations of motion is, in fact, a generic feature of
theories with Poincar\'e invariance. It is straightforward to show that
the geodesic equation of the physical metric is derived from
Eqs.~(\ref{N}-\ref{q}) when one chooses the unitary gauge, $q^a=0$. This
confirms the stated equivalence between the two formulations.

Classical solutions for the system may be obtained conveniently by choosing
a gauge, but  note,
first, that Eq.~(\ref{a}) requires
$\eta_3$ to be constant and we fix its value to be $\lambda$ to get
agreement with the geometric description in (\ref{eq1}).
Upon choosing the gauge $\eta^a =0$, a unique solution of Eq.~(\ref{ea}) is
%
%
%
\begin{equation}
\label{33}
e^a_\mu(x)=-\frac{4\pi G}{\lambda} p^a(\xi)\epsilon_{\mu\nu}
{\dot X^\nu(\xi)\over \dot X^0(\xi)}
\delta(\sigma-X^1(\xi))=-\frac{2\pi G}{\lambda} p^a(\xi)
\partial_\mu \epsilon (\sigma-X^1(\xi))
\label{zwei}
\end{equation}
where $\sigma \equiv x^1$, $t\equiv x^0$, $\xi(t)\equiv (X^0)^{-1}(t)$ and
$\epsilon(x)$ denotes the sign function of $x$.
Note also that in this gauge, Eq.~(\ref{omega}) becomes
%
%
%
\begin{equation}
\label{GHJ}
\partial_\mu \eta_2
=2 \pi G \epsilon_{ab}q^a(\xi) p^b (\xi)\partial_\mu \epsilon
\bigl(\sigma-X^1(\xi)\bigr),
\end{equation}
The quantity $\epsilon_{ab}q^a p^b$ in the right side of
the above equation is conserved in this gauge\footnote{\hsize=6.0truein
In an arbitrary  gauge one can show
that the gauge invariant combination
$B=\epsilon_{ab}(q^a+\frac{\eta^a(X(\tau))}{\eta^3})p^b$ is conserved using
particle equations of motion.
The existence of this conservation law is related to the invariance
under the Lorentz group. In the case of many particles, the conserved
quantity is $\sum_i B_i$.} and we call  its constant value $\nu$.
 The solution of Eq.~(\ref{GHJ}) reads
\begin{equation}
\label{tyt}
\eta_2(t,\sigma)=
{M\over 2\lambda}+ 2 \nu\epsilon (\sigma-X^1(\xi)).
\end{equation}
where $M$ is a new integration constant. It can be shown that
Eq.~(\ref{etaa}) is equivalent to Eq.~(\ref{q}) on the solution in
(\ref{zwei})
and that Eq.~(\ref{eta3}) simply fixes the form of the potential $a_\mu$
which does not  play a role in the following.
%
The  remaining  equation in
(\ref{eta2}) is  solved by
\begin{equation}
\label{36}
\omega_0
= -\frac{\pi G}{2}  p^2(\xi){N(\xi)\over \dot X^0(\xi)}
\epsilon(\sigma-X^1(\xi))\,,
\label{omega1}
\end{equation}
where we choose $\omega_1=0$ by using the
residual invariance under the Lorentz
transformations.

All the equations of motion for the particle on the solutions in
(\ref{GHJ}), (\ref{tyt}) and (\ref{omega1}) are reduced to
\begin{eqnarray}
\label{1para}
&&  \frac{
M+\lambda^2 q^2}{4\lambda}p^2 +m^2=0\ \ \ \ \ \ \ \ \ \ \ \\
&&\dot q^a -\frac{N }{4\lambda}p^a(M+
\lambda^2 q^2)=0\ \ \ \ \ \ \ \ \ \ \ \\
&&\dot p^a -\frac{N\lambda }{4}p^2q^a =0 ,\ \ \ \ \ \ \ \ \ \ \
\label{1parb}
\end{eqnarray}
where the prescription, $\epsilon(0)=0$ is used to specify $\eta_2(X)$
and $\omega_0(X)$. This excludes a possible self-interaction of the particle.

One striking fact at this point is that there is no equation for
the position variable, $X^\mu$, hence it is an arbitrary function
of time. The only requirement is that $\dot X^0(\tau)$ is positive
definite, which is necessary for the existence of
$\xi(t)$ in (\ref{33}--\ref{36}).

In the  one-body problem, in order to reach the unitary gauge we first
choose $X^a$ to be the same as $\epsilon^a_{~b} q^b$ by using the arbitrariness
of $X^a$, then we perform the gauge transformation generated by
\begin{eqnarray}
\theta^a(x)=-x^a-{2\pi G\over\lambda}p^a\epsilon(\sigma-X^1),\  \alpha(x)=0,
\label{gauget2}
\end{eqnarray}
where the coefficient of the sign function is determined such that the
transformed Zweibein does not involve any discontinuities and delta
function singularities.

We note that  (\ref{1para}-\ref{1parb}) with the
relation  $X^a=\epsilon^a_{~b}q^b$, are the first-order formulation of
the geodesic equation for a metric tensor
\begin{equation}
\label{metric10}
g_{ab}=\frac{4 h_{ab}}{{M\over \lambda}-\lambda X_c X^c},
\end{equation}
which describes the two dimensional black-hole.
The metric  also agrees with
the physical metric tensor felt at the particle position, namely
$\bar g^P_{\mu\nu}(X)=2\bar e^a_\mu(X) \bar e_{\nu a}(X)/\eta_2(X)$. The
geodesics, solutions of Eqs.~(\ref{1para}-\ref{1parb}), are given by
\begin{eqnarray}
X^0=-q^1={M^{1\over 2}\over 2\gamma\lambda}
\Bigl[e^{\nu \tau}(\gamma\cosh(\gamma\tau)-
\nu \sinh(\kappa\tau) )+ e^{-\nu \tau}(\gamma \cosh(\kappa\tau)+
\nu \sinh(\gamma\tau) )\Bigr],\\
X^1=-q^0={M^{1\over 2}\over 2\gamma\lambda}
 \Bigl[e^{\nu \tau}(\gamma \cosh(\gamma\tau)-
\nu \sinh(\gamma\tau) )-e^{-\nu \tau}(\gamma \cosh(\gamma\tau)+
\nu \sinh(\gamma\tau) )\Bigr]\,,
\end{eqnarray}
where
$\gamma\equiv \bigl(\nu^2-{4m^2}/{\lambda}\bigl)^{{1\over 2}}$.
The most general solutions are those obtained by applying a
$\tau$-independent Lorentz
transformations to the above solution.
The integration constant $M$ that is usually interpreted as a black hole mass,
should be positive for the reality of the solution. The cosmological
constant, $\lambda$,  should also be positive to avoid a geodesic solution
that describes a particle running into a naked singularity.
%
For real $\gamma$ ({\it i.e.} $\nu^2\ge 4m^2/\lambda$),
the geodesic represents a particle coming from  infinity and falling into
the black hole. On the other hand, the solution with imaginary $\gamma$
({\it i.e.} $\nu^2\ge 4m^2/\lambda$) corresponds to
a particle that comes out of the white hole and falls into the black hole.
%
%

Let us now turn to a many-body problem, described by an action
\begin{eqnarray}
\label{actionm}
I_N&=&\frac{1}{4\pi G}\int d^2 x \epsilon^{\mu\nu}\Bigl(\eta_a D_\mu e^a_\nu+
\eta_2 \partial_\mu\omega_\nu+\eta_3 (\partial_\mu a_\nu+\frac{1}{2}
\epsilon_{ab} e^a_\mu e^b_\nu)\Bigr)\nonumber\\
&+&\sum_i \int d\tau \Bigl [ p_{ia} (D_\tau q_i)^a-\frac{1}{2} N_i
\Bigl(\frac{q_i^A\eta_A(X_i) }{2}p_i^2
+m^2\Bigr) \Bigr ]
\end{eqnarray}
where $i$ labels the $i$-th particle and $X_i$, $q^a_i$ and $p^a_i$ are,
respectively, the position, Poincar\'e coordinates and Poincar\'e momenta
of the $i$-th particle. The equations of motion for the gravity sector
take the same form as Eqs.~(\ref{etaa}-\ref{a}) except that the right sides
of the equations are summed over all particles. The equations of motion
for the $i$-th particle obtained from variations with respect to $\delta N_i$,
$\delta q_i$ and $\delta p_i$ also take the same form
as those for the one-particle in (\ref{N}-\ref{q}). However, contrary
to the one-body problem, the $\delta X^\mu_i$ variation gives
linearly-independent
equations of motion, which may be reduced to  conditions
%
\begin{equation}
\label{pg}
(q^a_i-q^a_j)\delta^2( X_i-X_j)=0\,,
\end{equation}
when one uses the other equations. Eq. (\ref{pg}) simply
means that $q_i=q_j$ when $X^\mu_i=X^\mu_j$.

With the gauge choice $\eta_a(x)=0$ and $\omega_1=0$,
solutions
for the gravity sector read
\begin{eqnarray}
&&e^a_\mu=-\frac{2\pi G}{\lambda}\sum_i  p_i^a (\xi_i)
\partial_\mu \epsilon (\sigma-X^1_i(\xi_i))\\
&&\eta_2
={M\over 2\lambda}+ \sum_i \epsilon_{ab}q_i^a(\xi_i) p_i^b(\xi_i)\epsilon
(\sigma-X^1_i(\xi_i))\\
&&\omega_0
= -\frac{\pi G}{2} \sum_i  p_i^2(\xi_i){N_i(\xi_i)\over \dot X^0
(\xi_i)}\epsilon(\sigma-X^1_i(\xi_i)).
\end{eqnarray}
where $\xi_i=(X^0_i)^{-1}(t)$.
Taking these solutions into account, it is straightforward to show
that  all the remaining equations for the particles
are derived  from an effective Lagrangian,
\begin{eqnarray}
L=\sum_i \Bigl( p_{ia}\dot q_{i}^a -
{N_i\over 2}(\varphi_i p^2_i+m^2)\Bigr)
-\frac{4\pi G}{\lambda}\sum_{i < j}
\epsilon_{ab}p^a_i p^b_j
{d\over d \tau}\epsilon\bigl(X^1_i-X^1_j\bigr)
\label{lag10}
\end{eqnarray}
where
\begin{eqnarray}
\varphi_i={M+\lambda q^2_i\over 2\lambda}+\frac{2\pi G}{\lambda}
\sum_{i\ne j} \epsilon_{ab} q^a_j p^a_j
\epsilon\bigl(X^1_i-X^1_j\bigr)
\end{eqnarray}
and all the dynamical variables ($X_i,  q_i, p_i$ and $N_i$) are
functions of $\tau$. Note that the symplectic structure of this first
order Lagrangian is not in a canonical form. To achieve this, we
introduce a momentum  $\Pi_{i\mu}$ conjugate to $X^\mu_i$ and a new
set of constraints
\begin{equation}
M_i =\Pi_{i1}-\frac{4\pi G}{\lambda}\sum_{i\ne j}\epsilon_{ab}p^a_i p^b_j
\delta(X^1_i-X^1_j).
\end{equation}
Now the new Lagrangian reads
\begin{equation}
L=\sum_i \left ( p_{ia}\dot q_{i}^a+\Pi_{i\mu}\dot X^\mu_i -
{N_i\over 2}(\varphi_i p^2_i+m^2)-U_i M_i\right),
\label{lag90}
\end{equation}
where   $U_i$ are the Lagrange multipliers that enforce the new constraints.
It is interesting to notice that in the quantum case the particles will be
decribed by the same first order Lagrangian, once we have solved
gauge constraints.

As in the one-body case, the equations of motion do not determine
the particle trajectories $X^\mu_i$.
In addition, the equations of motion for $p^a_i$ and $q^a_i$
contain terms proportional to $\delta(X^1_i-X^1_j)$, which  implies that the
Poincar\'e coordinates $q_i^a$ must be discontinuos when particles meet
({\it i.e.} $X_i=X_j$). Therefore, the relation
between  the Poincar\'e coordinates and the
particle trajectories is  not as simple as in the  one particle case.

All of these troubles disappear when one performs a gauge transformation
generated by
\begin{eqnarray}
\theta^a(x)=\!-x^a\!-\!{2\pi G\over\lambda}\sum_i p_i^a\epsilon(\sigma-X^1_i),\
\alpha=0,
\label{gauget3}
\end{eqnarray}
with a diffeomorphism gauge choice
\begin{eqnarray}
X_i^a=\epsilon^a_{~b} q_i^b+{2\pi G\over\lambda}\sum_{j\ne i} p_j^a
\epsilon(X^1_i-X^1_j)\,.
\label{iden3}
\end{eqnarray}
As in the one-body case, this transformation leads to the unitary gauge.
In (\ref{gauget3}), the coefficients of the sign function are chosen
such that the transformed Zweibein does not involve  any
singularities and discontinuities. The identification in (\ref{iden3})
is obtained simply from the invariance of the quantities
$q^a_i+{\eta(X_i)/\lambda}$ under the above transformation.

The sign function in (\ref{iden3})  removes discontinuous
parts from the Poincar\'e coordinates, so that $X^a_i$ describes a
continuous particle trajectory. Moreover, with the identification
in (\ref{iden3}), the conditions in (\ref{pg}) are trivially satisfied.

In this unitary gauge, using Eq.~(\ref{iden3})
we can rewrite  the effective Lagrangian  (\ref{lag10})
in terms of $X^a_i$ and $P^a_i\equiv \epsilon^a_{~b} p^b_i$
\begin{eqnarray}
L=\sum_i \Bigl( P_{ia}\dot X_{i}^a -
{N_i\over 2}(\bar\varphi_i P^2_i-m^2)\Bigr)
\label{lag100}
\end{eqnarray}
where
\begin{eqnarray}
\bar\varphi_i&=&{M\over 2\lambda}-
{\lambda\over 2}\Bigl (X^a_i-{2\pi G\over \lambda}\sum_{j\ne i}
\epsilon^a_{~b} P^b_i\epsilon(X^1_i-X^1_j)\Bigr)
(X_{ia}-{2\pi G\over \lambda}\sum_{j\ne i}
\epsilon_{ac} P^c_i\epsilon(X^1_i-X^1_j)\Bigr)\nonumber\\
&-&\!\!\!\!\!\frac{2\pi G}{\lambda}
\sum_{i\ne j} \epsilon_{ab} \Bigl(X^a_j\!-\!{2\pi G\over\lambda}
\!\sum_{k\ne j}\!\epsilon^a_{~c} P^c_k\epsilon(X^1_j\!-\!X^1_k)\Bigr) P^b_j
\epsilon\bigl(X^1_i-X^1_j\bigr)\,.
\label{nbody}
\end{eqnarray}
We present here   the Lagrangian in a
reparametrization invariant form, so it should be noted that
all the dynamical variables
in the Lagrangian are functions of $\tau$.

This  Lagrangian  does not include any delta terms and, therefore, the
problem arising from the delta terms in $(\ref{lag10})$  disappears.
It  provides us with a clear picture of the system where all the
gauge degrees of freedom  have been eliminated. The geometry is
now hidden in the factor $\phi_i$. In fact, the action for each particle
can be considered  as the geodesic action for the metric $h_{ab}/\phi_i$.

\typeout{fine}
\newsection{Quantization}\indent

\noindent{\bf One-body problem}

We first quantize the one-body  problem.
It turns out that for this case, one can consistently solve the constraints
to find the most general
wave-functional.

Owing to the gauge symmetry, the Lagrangian contains
Gauss law constraints, which we proceed to solve first in
order to find the most general
gauge-invariant wave functional. Since the theory possesses also
a general coordinate invariance, we encounter a Hamiltonian that consists
of constraints only.  Thus the
quantization involves solely solving the constraints, by which one may
find physical wave functionals.

Let us begin with recording the first order Lagrange density with
a reparametrization gauge, $t\equiv x^0= X^0(\tau)$,  which  reads
\begin{eqnarray}
{\cal L}&=&
{1\over \ggg}(\eta_a{\dot e}^a_1+\eta_2{\dot \omega}_1+ \eta_3{\dot a}_1)
+e^a_0 G_a + \omega_o G_2 + a_0 G_3 \nonumber \\
 &+&\Bigl( p_a {\dot q}^a +p^a \epsilon_{ab}
\Bigl(q^b \omega_1(X) -e^b_1(X)\Bigr){\dot X^1}
-
N ({1\over 2}\eta_A q^A p^2 +m^2)\Bigr) \delta (\sigma -X^1(\tau))
\label{laga}
\end{eqnarray}
where
dot/dash signifies
time/space derivatives. The Gauss generators are
\begin{eqnarray}
 G_a(\sigma)&=&
{1\over \ggg}(\eta'_a+\epsilon_{ab}\eta^b\omega_1+ \eta_3
\epsilon_{ab}e_1^b)+\epsilon_{ab}p^b \delta (\sigma -X^1(\tau))\\
  G_2(\sigma)&=&
{1\over \ggg}(\eta'_2+\eta^a\epsilon_{ab}e^b_1)
-q^a\epsilon_{ab}p^b \delta (\sigma -X^1(\tau)) \\
G_3(\sigma) &=&{1\over \ggg}\eta'_3 \ .
\label{gauss}
\end{eqnarray}
Since the symplectic structure for the  dynamical variable $X^1$
is not in a standard form, we introduce one more constraint,
\begin{eqnarray}
 M(X)\equiv \Pi- p^a \epsilon_{ab} \Bigl(q^b \omega_1(X) -e^b_1(X)\Bigr)
\label{mome}
\end{eqnarray}
with help of a Lagrange multiplier $U$.
Thus, the Lagrangian is equivalently presented as
\begin{eqnarray}
 {\cal L}&=&
{1\over \ggg}(\eta_a{\dot e}^a_1+\eta_2{\dot \omega}_1+ \eta_3{\dot a}_1)
+ ( p_a {\dot q}^a +\Pi {\dot X^1})\delta(\sigma -X^1)\nonumber \\
 &+&e^a_0 G_a + \omega_o G_2 + a_0 G_3
- \{N H(X) + U M(X)\}\delta(\sigma -X^1)
\label{lagb}
\end{eqnarray}
where $H(X)$ denotes the mass-shell constraint ${1\over 2}\eta^A(X) q_A
p^2 +m^2$
in the background geometry ${2\eta_{\mu\nu}/ \eta_A q^A}$. The terms
in the first line of the Lagrangian describe the symplectic structure of
the theory, while the remaining terms work as constraints. When
implementing the Dirac procedure, we first
need to check  whether there are any secondary constraints by
commuting primary constraints with one another. The Gauss law generators
simply produce the Lie algebra of the $ISO(1,1)$ group. The algebra
is
\begin{eqnarray}
 &\ &\!\!\!\!\!\!\!\!\!\!\!\!\!\!\!\!\!\!\!\!
[G_a(\sigma), G_b(\sigma')] = i \epsilon_{ab}G_3(\sigma)
\delta(\sigma-\sigma') \\
&\ &\!\!\!\!\!\!\!\!\!\!\!\!\!\!\!\!\!\!\!\!
[G_a(\sigma),G_2(\sigma')] = i \epsilon_{ab}G^b(\sigma)
\delta(\sigma-\sigma')  \\
&\ &\!\!\!\!\!\!\!\!\!\!\!\!\!\!\!\!\!\!\!\!
[G_A(\sigma),G_3(\sigma')] = 0
\label{coma}
\end{eqnarray}
In addition, the momentum and mass-shell constraints are gauge invariant,
{\it i.e.}
\begin{eqnarray}
 [G_A(\sigma),H(X)] = [G_A(\sigma),M(X)]= 0
\label{comb}
\end{eqnarray}
and finally the commutation between mass-shell  and
momentum constraints is
\begin{eqnarray}
 [M(X),H(X)] = -i \gggg q^A p^2 G_A(X)\ .
\label{comc}
\end{eqnarray}
Hence, the algebra of the constraints is closed, so the constraints are all
first--class.
(When we discuss two-particle problem, the closure  of the constraint algebra
will be at issue again, and in fact it will be shown that the algebra does not
close.)

For each dynamical variable, we may use either coordinate representation
or momentum representation: the two are related by an appropriate
Fourier transform. In our case, we shall use momentum representations for
gravitational and Poincarr\`e variables, $A_1^A \equiv (e^a_1, \omega_1,a_1)
=(i\ggg{\delta/\delta\eta_a},i\ggg{\delta/\delta\eta_2},
i\ggg{\delta/\delta\eta_3})$,
$q_a=i{\ppp/\ppp p^a}$, while we shall use the
usual coordinate representation, $\Pi={\ppp/ i\ppp X^1}$
for the particle position, $X^1$.
Accordingly  the wave function or functional is
a function or functional of $\eta^A, p^a$ and $X^1$.

As in \cite{Can93}, the most general solution of the gauge constraints
in the momentum representation is
\begin{eqnarray}
 \Phi=e^{i(\Omega(\eta) +{\ggg\over \lambda} p^a\eta_a(X))}\delta(\eta'_3)
\delta((\eta^A\eta_A)'-\gggg\delta(\sigma-X) q^a\epsilon_{ab}p^b)
\Psi(p^a, M, \lambda, X^1)
\label{sola}
\end{eqnarray}
where  $\Omega$ is an integral of the Kirillov--Kostant one--form
\begin{eqnarray}
 \Omega = \int \eta^a\epsilon_{ab} d\eta^b/\eta_3 \ ,
\label{kiri}
\end{eqnarray}
and $M$ and $\lambda$ are, respectively, constant parts of $
\eta^A\eta_A$ and $\eta_3$. Note that the second delta function
in (\ref{sola}) includes an operator $q^a\epsilon_{ab} p^b$, which
could be diagonalized if $\psi$ is an eigenfunction see below.

As is   usual in gauge theory, imposing the  Gauss law constraint
on the wave functional ensures gauge invariance
of the ``coordinate'' representation wave functional.
However, in the ``momentum'' representation, one expects that the wave
functional is  gauge-invariant up to a phase factor. This may be
understood as follows. The momentum representation is related with the
coordinate representation by a functional Fourier transform,
$\Phi(\eta)=\int DA e^{-i\int dx A^A \eta_A} {\bar\Phi}(A)$.
Since the measure $DA$ and ${\bar \Phi}(A)$ are gauge invariant but
$\int dx A^A \eta_A$ is not, the wave functional in the momentum
representation is not gauge invariant. In fact, it has a structure of gauge
invariant functional multiplied by phase exponential of the integrated
Kirillov--Kostant one--form, which is denoted by $\Omega$ above.

When one applies to $\Phi$ the remaining constraints, the mass-shell and the
momentum,  one obtains new constraints on
$\Psi(p^a, M, \lambda, X^1)$.
The new momentum constraint reads
\begin{eqnarray}
\Pi \Psi = {\ppp\over i\ppp X^1} \Psi =0\ ,
\label{momb}
\end{eqnarray}
which implies that the wave function does not depend on the particle
coordinate $X^1$. On the other hand, the mass-shell
constraint takes the form,
\begin{eqnarray}
\Bigl[ {\lambda\over 4} (q^2 + {M\over \lambda^2})p^2 +m^2 \Bigr]
\Psi(p^a) =0\ .
\label{masb}
\end{eqnarray}
The equation may be interpreted as describing a particle moving in a
background geometry with a metric,
\begin{eqnarray}
{\bar g}_{\mu\nu}= {4\eta_{\mu\nu}\over \lambda(q^2+ M/\lambda^2)}\ .
\label{meta}
\end{eqnarray}
The metric
${\bar g}_{\mu\nu}$ is not affected by the particle state since its form
is determined by the contribution from the pure gravity sector. Thus,
there are
essentially no self-interactions.

For the ordering convention of the noncommuting operators,
$p^a$ and $q^a$, we follow the same ordering as that in the
Laplace-Beltrami operator with a metric ${\bar g}_{\mu\nu}$.
With this ordering, when one takes the $m=0$ limit, the constraint
is reduced to $p^2 \Psi(p) =0$. This is the expected result because
for $m=0$, the theory possesses a conformal invariance and the metric
${\bar g}_{\mu\nu}$ is conformally flat.

Since the boost generator $q^a\epsilon_{ab} p^b$ commutes with the
mass-shell constraint, we diagonalize it:
\begin{eqnarray}
q^a\epsilon_{ab} p^b \Bigl({p^0+p^1\over p^0-p^1}\Bigr)^{{i\nu\over 2}}=
\nu  \Bigl({p^0+p^1\over p^0-p^1}\Bigr)^{{i\nu\over 2}}\ ,
\label{boos}
\end{eqnarray}
where $\nu$ is a real eigenvalue of the boost generator ranging from
$-\infty$ to $\infty$.

Finally, the solution of the mass-shell constraint is given by
\begin{eqnarray}
\Psi (p^a)=\Bigl({p^0+p^1\over p^0-p^1}\Bigr)^{{i\nu\over 2}}
p^2\psi_\nu(-p^2)\ ,
\label{wave}
\end{eqnarray}
where $\psi_\nu (z)$ satisfies the differential equation:
\begin{eqnarray}
\Bigl[ {d^2\over dz^2} + {d\over zdz} + {1\over z^2}
({\nu^2\over 4}- {m^2\over \lambda}) - {M\over 4\lambda^2 z}\Bigr]
\psi_\nu (z)\ .
\label{bess}
\end{eqnarray}
This  is solved by the Bessel function
\begin{eqnarray}
 \psi_\nu(z)=Z_{i\gamma}(\beta z^{1\over 2})\ ,
\label{defa}
\end{eqnarray}
where $\gamma =\sqrt{ \nu^2-{4m^2\over \lambda}}$, $\beta=
{M^{{1\over 2}}\over \lambda}$ and $Z_{i\gamma} (x)$ denotes the Bessel
function of imaginary argument $K_{i\gamma}(x)$ or $I_{i\gamma}(x)$.
(Here, we assume that the cosmological constant $\lambda$
and the black hole mass $M$
are
positive.)


Note that the asymptotic behavior of the Bessel functions
are
\begin{eqnarray}
 I_{i\gamma} (z)\rightarrow {e^z\over \sqrt{2\pi z}}\ , \ \
 K_{i\gamma} (z)\rightarrow {e^{-z}\over \sqrt{2\pi z}}\ .
\label{asym}
\end{eqnarray}
Therefore, one has to exclude $I_{i\gamma}$, since a wave function in
the position space for $I_{i\gamma}$ defined  by a Fourier transform
 does not exist.

The meaning of this one-body wave function is clearly seen from the
Poincar\'e coordinate space. The Fourier transform of the wave function
with
$e^{iq^a p_a}$ gives a coordinate-space wave function  depending solely
on the combination $\rho^a(X)=q^a +{\eta^a(X)\over \lambda}$ [cf.
 (\ref{sola})]. Thus,
in the unitary gauge $q^a=0$, choosing the diffeomorphism gauge
${\eta^a(X)\over \lambda}=\epsilon^a_{~b}X^b$, we obtain a usual
interpretation in terms of the position variable $X$. It is a
wave function describing a particle moving in the  geometry characterized
by the metric in (\ref{metric10}).


\noindent {\bf Two-Body problem}

Let us now turn to the many-body problem, which is more involved than
the one-particle case since there are interactions between particles
through the coupling of gravity.
For simplicity, we concentrate on the two-body problem in the following.
The strategy of solving
the constraints are the same as in the one-particle problem.

The gauge constraints read
\begin{eqnarray}
 G_a(\sigma)&=&
{1\over \ggg}(\eta'_a+\epsilon_{ab}\eta^b\omega_1+ \eta_3
\epsilon_{ab}e_1^b)+\sum_i\epsilon_{ab}p_i^b \delta
(\sigma -X^1_i(\tau))\, , \\
  G_2(\sigma)&=&
{1\over \ggg}(\eta'_2+\eta^a\epsilon_{ab}e^b_1)
-\sum_i q_i^a\epsilon_{ab}p_i^b \delta (\sigma -X^1_i(\tau))\, ,\\
G_3(\sigma) &=&{1\over \ggg}\eta'_3 \ .
\label{gausb}
\end{eqnarray}
As one sees from the above, the Gauss law constraint consists of a sum of
the contributions from each particle. Mass-shell and momentum constraints
are, respectively,
\begin{eqnarray}
 M_i(X_i)&\equiv& \Pi_i- p_i^a \epsilon_{ab} (q_i^b \omega_1 -e^b_1)\, ,\\
 H_i(X_i)&\equiv& {1\over 2}\eta_A q_i^A p_i^2 +m^2\,.
\label{momd}
\end{eqnarray}
As in the one-particle problem, the Gauss law generators, $G^A$ satisfies
Lie algebra in (\ref{coma}), and $M_i(X_i)$ and $H_i(X_i)$ are also
gauge invariant: $M_i(X_i)$ and $H_i(X_i)$ commute with the Gauss
law generators and $[M_1, M_2]=[H_1,H_2]=0$. The
remaining commutation relations are
\begin{eqnarray}
[M_1+M_2, H_i]  &=& -i\gggg q_i^A p_i^2 G_A(X_i)\ , \\
\left[M_1- M_2, H_1\right] &=& -i\gggg q_1^A p_1^2 G_A(X_1)
+i\ggg (q_1-q_2)^a\epsilon_{ab} p_2^b \delta (X^1_1-X^1_2)p_1^2\ , \\
\left[M_1-M_2, H_2\right] &=& i\gggg q_2^A p_2^2 G_A(X_2)
+i\ggg (q_1-q_2)^a\epsilon_{ab} p_1^b \delta (X^1_1-X^1_2)p_2^2\ .
\label{comd}
\end{eqnarray}
where the fact that $p_i^2$ commutes with $G_A$, has been used.
Thus the constraint algebra is not closed unless $X^1_1$ is different from
$X^1_2$. One might think that this problem is restricted to the zero measure
 subset of the whole phase space, so that one may ignore the problem,
for example, by an appropriate boundary condition\cite{kab}.
However, this cannot be
achieved consistently in this case, as we shall see in the following.

At this point, one may try to find all possible secondary constraints
by commuting repeatedly the induced constraint with
the primary constraints. However, we shall not 	determine
the secondary constraints, rather  we show from just
the primary constrains, that the theory does not admit any solutions.

Let us begin by solving the gauge constraints, where the
most general solution is
\begin{eqnarray}
 \Phi=e^{i(\Omega(\eta) +{\sum_i}{\ggg\over\lambda}
p_i^a\eta_a(X_i))}\delta(\eta'_3)
\delta\{(\eta^A\eta_A)'\!\!-\!\!\sum_i\ggg\lambda
\delta(\sigma\!\!-\!\!X^1_i) q_i^a\epsilon_{ab}p_i^b\}
\Psi(p_i^a, M, \lambda, X^1_i)
\label{sola1}
\end{eqnarray}
Operating the mass-shell constraints and the momentum constraints on
(\ref{sola1}), we obtain simplified version of the constraints on $\Psi$,
\begin{eqnarray}
{\bar M}_1 &=& \Pi_1 - {\ggg\over \lambda}p^a_1\epsilon_{ab} p^b_2
\delta(X^1_1-X^1_2) \ , \nonumber \\
{\bar M}_2 &=& \Pi_2 + {\ggg\over \lambda}p^a_1\epsilon_{ab} p^b_2
\delta(X^1_1-X^1_2) \ , \nonumber \\
{\bar H}_1&=&{\lambda\over 4}\Bigl(q_1^2 + {M\over \lambda^2}
+{\ggg\over\lambda} q_2^a\epsilon_{ab} p^b_2\epsilon(X^1_1-X^1_2)\Bigr)
p_1^2 +m^2 \ ,
\nonumber\\
{\bar H}_2&=&{\lambda\over 4} \Bigl(q_2^2 + {M\over \lambda^2}
-{\ggg\over\lambda} q_1^a\epsilon_{ab} p^b_1\epsilon(X^1_1-X^1_2)\Bigr)
p_2^2 +m^2 \ .
\label{masc}
\end{eqnarray}
Note that this set of constraint can be obtained directly from the effective
Lagrangian (\ref{lag90}) for the case of two particles.
As is  seen in the mass-shell constraints, the particles interact via
gravitation: for example, the metric of the first particle
depends on the motion of the second particle. It is found that analyzing
the constraints in a new coordinate system, $X_+={X^1_1 +X^1_2\over 2}$
and $X_-=X^1_1-X^1_2$, is particularly convenient. Accordingly, we define
$M_+\equiv M_1+M_2 ={\ppp/i\ppp X_+}$ and $M_-\equiv {1\over 2}
(M_1-M_2) ={\ppp/i\ppp X_-}-{\ggg\over\lambda} \epsilon_{ab}
p_1^ap_2^b\delta(X_-)$.
The total momentum constraint, $M_+$ commutes with all the other
constraints, and  merely implies that the wave function $\Psi$
does not depend on the coordinate $X_+$.

The remaining three require more careful analysis. Above all, the
expressions involve $\delta(x)$ and the sign function $\epsilon (x)$,
which need to be regularized to deal with, for example, $\delta(x)
\epsilon^2(x)$ at $x= 0$.
First, we take a following regularization defining
\begin{eqnarray}
\epsilon_\Lambda (x)&\equiv& \tanh{\Lambda x}\ , \nonumber\\
\delta_\Lambda(x)&\equiv&{\phantom{2}d\over 2dx}\epsilon_\Lambda (x)\ .
\label{regu}
\end{eqnarray}
To get a representation of $\epsilon(x)$, we take the limit in which $\Lambda$
goes to infinity. Of course, during a specific computation, we do the
computation with a generic $\Lambda$ and in the end take the
$\Lambda\rightarrow \infty$
limit.

With the above regularization, we have the following rules of computation:
\begin{eqnarray}
{d\over dx}\epsilon^n_\Lambda (x)&=& n \epsilon^{n-1}_\Lambda (x)
\delta_\Lambda(x)\ , \\
\lim_{\Lambda\rightarrow\infty}\int \delta_\Lambda(x)\epsilon^n_\Lambda
(x)f(x)&=&
\lim_{\Lambda\rightarrow\infty}\int {1\over 2(n+1)}{d\over
dx}\epsilon^{n+1}_\Lambda
(x)f(x)dx\nonumber\\
&=&
-\lim_{\Lambda\rightarrow\infty}\int {1\over 2(n+1)}\epsilon^{n+1}_\Lambda(x)
{d\over dx}f(x)dx\nonumber\\
&=& {1\over 2(n+1)}\left[f(0^+) - (-1)^{n+1}f(0^-)\right]\nonumber\\
&=&\left[
\begin{array}{ll}
0 & n= {\rm odd}\\
{f(0)\over n+1} &  n= {\rm even}
\end{array}\right.
\label{regb}
\end{eqnarray}
where the continuity of the test function $f(x)$ is assumed and $n > 0$.
Hence,
\begin{eqnarray}
\delta (x)\epsilon^n(x)=\left[
\begin{array}{ll}
0 & n= {\rm odd}\\
{\delta(x)\over n+1}
&  n= {\rm even}
\end{array}\right.
\label{regc}
\end{eqnarray}
Obviously one  has to compute with  arbitrary $\Lambda$ and then
take the limit in the end. We shall follow  this way to solve our problem.

The regulated equations one has to solve to find the two-body wave function
are
\begin{eqnarray}
{\hat M}\psi &=&\Bigl[ {\phantom{1}\ppp\over i\ppp X_-}
- {\ggg\over \lambda}\epsilon_{ab}p^a_1p^b_2
\delta_\Lambda(X_-)\Bigr]\psi \ ,  \\
{\hat H}_1\psi&=&\Bigl[ \Bigl(q_1^2 - {M\over \lambda^2}
+{\ggg\over\lambda}\epsilon_{ab} q_2^a p_2^b\epsilon_\Lambda(X_-)\Bigr)
 +{4m^2\over \lambda p^2_1} \Bigr]\psi\ , \\
{\hat H}_2\psi&=&\Bigl[ \Bigl(q_2^2 - {M\over \lambda^2}
-{\ggg\over\lambda}\epsilon_{ab} q_1^a p_1^b\epsilon_\Lambda(X_-)\Bigr)
 +{4m^2\over \lambda p^2_2} \Bigr]\psi\ .
\label{masd}
\end{eqnarray}
where $\psi(p^a_1,p^a_2)=p^2_1p^2_2\Psi (p^a_1,p^a_2)$.

The commutation relations between ${\hat M}$ and ${\hat H}_1$/${\hat H}_2$
are now given by
\begin{eqnarray}
\label{come1}
[{\hat M},{\hat H}_1 ]  &=& i{8\pi G\over \lambda}
 \Bigl[\epsilon_{ab}(q_1-q_2)^a p_2^b \delta_\Lambda (X_-) +
{\gggg\over\lambda}\delta_\Lambda(X_-)
\epsilon_\Lambda(X_-)p_1\cdot p_2 \Bigr]\ ,\\
\left[{\hat M},{\hat H}_2\right ]  &=& i{8\pi G\over \lambda}
 \Bigl[\epsilon_{ab}(q_1-q_2)^ap_1^b
\delta_\Lambda (X_-) -{\gggg\over\lambda}\delta_\Lambda(X_-)
\epsilon_\Lambda(X_-)p_1\cdot p_2 \Bigr]\ .
\label{come}
\end{eqnarray}
Since $[{\hat M},{\hat H}_1 ]\psi=[{\hat M},{\hat H}_2 ]\psi=0$
and $\delta_\Lambda(x)$ and $\delta_\Lambda (x)\epsilon_\Lambda(x)$
are independent functions of $x$ for all $\Lambda$, we conclude that
\begin{eqnarray}
 \epsilon_{ab}(q_1-q_2)^ap_2^b \psi =
\epsilon_{ab} (q_1-q_2)^a p_1^b \psi =p_1\cdot p_2\psi=0 \ .
\label{cons}
\end{eqnarray}
The last equality implies that $\psi\ \propto \ \delta(p_1\cdot p_2)$ and
the first two  give that  $\epsilon_{ab}p_1^a p_2^b
\delta'(p_1\cdot p_2)=0$. The only possible solution,
$\psi\ \propto\  \delta(p_1)\delta(p_2)$, is excluded from the
consideration of $H_1\Psi=0$. Thus the solution of the constraints
does not exist.

When we solve ${\hat M}\psi={\hat H}_1\psi={\hat H}_2\psi=0$, we may use
another regularization scheme. For example, a regularization may be
consistently defined by $\delta(x)\epsilon(x)=0$, ${\phantom{1}d\over dx}
\epsilon(x)=2\delta(x)$ and
{\small $
\epsilon^n(x)=\left[
\begin{array}{ll}
\epsilon(x) & n={\rm odd}\\
\,\, 1           & n={\rm even}
\end{array}\right.$}.
In addition, all differentiation with respect to $x$ should be applied
after the third equation is used
completely. With this regularization, though
it is complicated, one may show again that a  solution of
the three constraints does not exist.

The obvious question  is why the theory does not lead to
any consistent solutions.
As shown in the classical analysis, equivalence between the geometric
approach and the gauge theoretic formulation provides the relations
$X^a=\epsilon^a_{~b}q^b$ and
(\ref{iden3}) for the $N$-body problem.
For the one-body case, this relation is used to reinterpret
the wave function.
For the two-body case, the wave function depends on both the
Poincar\'e coordinates and $X_i^\mu$. Consequently, without this relation,
equations for the Poincar\'e momentum and the position variables put
separately too much restrictions. [For example, the conditions in (\ref{pg})
corresponds to  new constraints in (\ref{come1}-\ref{come}) upon quantization,
which cause troubles in finding solutions,
whereas they become trivial once the  relation $X^a=\epsilon^a_{~b} q^b$
is used]. As a result, a consistent solution for the two-body problem
does not exist.

One obvious resolution of this problem is as follows. First, using the
relation (\ref{iden3}), identify the classical phase space before
quantization, which leads to the effective Lagrangian in (\ref{lag10}).
Next we quantize the system solving constraints, which is now perfectly
well defined and consistent.

Another possibility is to look for a reasonable modification of the
constraint that makes them compatible. For example one can consider
$\tilde M=X_- \hat M$ instead of $\hat M$.  The algebra that is formed
by $ \tilde M$,  $\hat H_1$ and $\hat H_2$ is now closed and solvable.
In fact using the result of the one body  problem, we can  easily show
that the most general solution is
\begin{eqnarray}
\psi &=&
\!\left(\frac{p^0_1+p_1^0}{p^0_1-p^1_1}\right)^{i \nu_1\over 2}\!\!\left(
\frac{p^0_2+p_2^0}{p^0_2-p^1_2}\right)^{i \nu_2\over 2}
\!\!\biggl [\Bigl(1+\epsilon(X_-)\Bigr)
K_{i\gamma_1}\Bigl(\frac{\sqrt{M+2\pi G \nu_2}}{\lambda}
\sqrt{\rho_1^a\rho_{1a}}\Bigr ) \\
&&\times K_{i\gamma_2}\Bigl (\frac{\sqrt{M-2\pi G \nu_1}}{\lambda}
\sqrt{\rho_2^a\rho_{2a}}\Bigr )+\Bigl(1-\epsilon(X_-)\Bigr)({\nu_1,\nu_2}\to
{-\nu_1,-\nu_2}) \biggr ]\nonumber
\end{eqnarray}
where $\gamma_i=\sqrt{\nu_i^2- { 4 m^2}/{\lambda}}$.
The change considered here corresponds to modifying the contact interaction
when
the particles meet. However, one should note that there is no clear reason
for such a modification.
\newsection{Conclusions}\indent

In this paper, we present the gauge theoretic model for the string-inspired
gravity with point particles. For a gauge  invariant description
of particles we enlarge the phase space by introducing the
Poincar\'e coordinate.
We also use a non-minimal gauge interaction with
$\eta_A$ fields in order to  make particles couple to the
physical metric.

In the investigation of the classical problem, we first show the equivalence
to the geometric approach by comparing the equations of motion in the unitary
gauge and obtain one-body solutions that describe a particle in the
black-hole metric.
For the many-body case, we derive the effective action
for the Poincar\'e coordinates and the particle positions in which
the gravity variables are  eliminated  completely.
Transforming the action into the unitary
gauge, we are led to the new effective action for the particle position
only, from which one may see clearly its geometrical implications.

After solving the wave functional of the gravity part, the quantum one-body
problem is reduced to the Klein-Gordon equation with the black hole metric,
which implies that the classical picture for the geometry is preserved at
the quantum level.

On the other hand, in the many-body case, the enlargement of the phase space
makes the constraint algebra open at the points where particle coordinates
coincide. Consequently, we prove that  it is impossible to solve
the constraints
consistently. Thus one may conclude that the classical
equivalence between the two
formulations is lost upon quantization.

A way of overcoming the problem is to reduce the phase space first by
adopting the unitary gauge and then to quantize the theory within
the reduced phase space. This procedure essentially leads to
the problem of quantizing the effective
 action in (\ref{lag100}) that provides a closed
constraint algebra and a well-defined symplectic structure. Detailed
quantum analysis of the action might be enlightening in understanding
the mutual effects of geometry, particles and gravity interaction.

A simpler problem may be an investigation of the one-body wave functional
near  the  singularity and the horizon, and effects of the particle state
to the geometry.

It is also interesting to pursue  a direct quantization of the geometric
model with point particles, so that, for example, one may see a natural
diffeomorphism gauge choice in the geometric theory by solving constraints.

\bigskip
\begin{center}
{\bf ACKNOWLEDGEMENTS}\end{center}
\ \indent
We would like to acknowledge Professor R. Jackiw for
enlightening discussions and careful reading of the manuscript.
We also thank G. Amelino-Camelia for helpful comments.
\hfill
\eject

\end{document}